\documentclass[a4paper]{jpconf}
\usepackage{graphicx}
\newcommand{\be}{\begin{equation}}
\newcommand{\ee}{\end{equation}}
\newcommand{\bea}{\begin{eqnarray}}
\newcommand{\eea}{\end{eqnarray}}
\newcommand{\nn}{\nonumber\\}
\newcommand{\beas}{\begin{eqnarray*}}
\newcommand{\eeas}{\end{eqnarray*}}

\newcommand{\slshh}[1]{{\not \!\! #1}}

\begin{document}

\title{Hadronic matter at the edge: A survey of some theoretical approaches to the physics of the QCD phase diagram} 
\author{Alejandro Ayala} 
\address{Instituto de Ciencias
  Nucleares, Universidad Nacional Aut\'onoma de M\'exico, Apartado
  Postal 70-543, M\'exico Distrito Federal 04510,
  Mexico\\ and\\
  Centre for Theoretical and Mathematical Physics, and Department of Physics,
  University of Cape Town, Rondebosch 7700, South Africa.}

\begin{abstract}

In the past few years a wealth of high quality data has made possible to test current theoretical ideas about the properties of hadrons subject to extreme conditions of density and temperature. The relativistic heavy-ion program carried out at the CERN-SPS and under development at the BNL-RHIC and CERN-LHC has provided results that probe the evolution of collisions of hadronic matter at high energies from the initially large density to the late dilute stages. In addition, QCD on the lattice has produced results complementing these findings with first principles calculations for observables in a regime where perturbative techniques cannot describe the nature of strongly coupled systems. This work aims to review some recent developments that make use of field theoretical methods to describe the physics of hadrons at finite temperature and density. I concentrate on two of the main topics that have been explored in the last few years: (1) The search for the structure of the phase diagram and (2) analytical signals linked to the chiral symmetry restoration/deconfinement. 

\end{abstract}

\section{The QCD phase diagram}\label{1}

The different phases in which matter made up of quarks and gluons arranges itself depends, as for any other substance, on temperature and density or equivalently, on the chemical potentials coupled to the corresponding quark quantum numbers. Our experimental knowledge on the structure of the phase diagram comes basically from information collected in relativistic heavy-ion collisions. A working theoretical assumption is that local equilibrium is a good description after the early stages of the collision. This assumption, together with the large amount of particles produced in the reactions, emphasizes the need to account for the density variable. Since $u$, $d$ and $s$ are the relevant quark species for temperatures and densities close to the phase boundaries, a complete description should in principle be given in terms of possibly distinct chemical potentials describing the density associated to these quarks. Nevertheless, in equilibrium, these chemical potentials are not independent from each other. There are two basic requirements that they need to satisfy: beta equilibrium and charge neutrality. This means that out of the three chemical potentials, we can choose one of them as independent. In this manner, the phase diagram becomes two dimensional. 

\begin{figure}[t!]
\centering\includegraphics[width=\columnwidth]{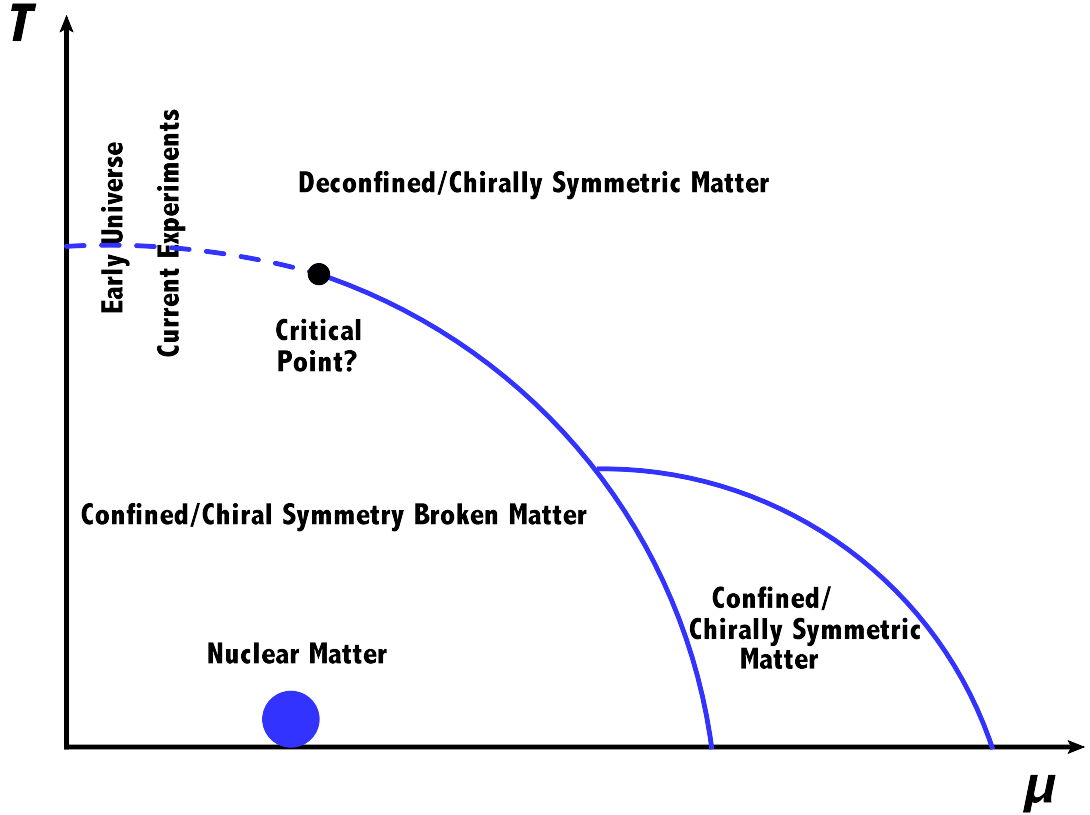}
\caption{Schematic representation fo the QCD phase diagram in the $T-\mu$ plane. }
\label{fig1}
\end{figure}

A popular representation of this two-dimensional phase diagram, where the independent chemical potential chosen is $\mu$, the light-quark chemical potential, is given in Fig.~\ref{fig1}. $\mu$ is related to the baryon chemical potential $\mu_B$ by $\mu_B=3\mu$. It should be noted that most of our knowledge of the phase diagram is restricted to the $T\geq 0$, $\mu=0$ axis. The phase diagram is by and large unknown. For physical quark masses and $\mu=0$ lattice calculations have shown~\cite{Aoki} that the change of phase, from where the dominant degrees of freedom are quarks and gluons (QGP), to where the dominant ones are hadrons, is an analytic crossover and thus it is not a real phase transition. This means that the change is smooth, as opposed to presenting a jump in some of the thermodynamic variables. This result was first obtained analytically in Ref.~\cite{Yukalov1} from a model of phase transitions for statistical systems that can possess different bound states, described as compact clusters (see also Ref.~\cite{Yukalov2}). Given that there is no true phase transition to the QGP as the temperature increases, the definition of the (pseudo) critical temperature, $T_c$, is to some extent a matter of choice. Since for systems where there is a true phase transition, the transition lines are defined in terms of the equality of the free energies in one phase and the other, one can resort to analyzing observables sensitive to the free-energy to define $T_c$. 

To describe chiral symmetry restoration, one of these observables is the quark condensate $\langle\bar{\psi}\psi\rangle$ which, from the Hellmann-Feynman theorem generalized to statistical systems, can be written as
\bea
\label{Hellmann-Feynman}
   \langle\bar{\psi}\psi\rangle=\frac{d}{dm}\Omega,
\eea
where $\Omega$ is the grand potential or free-energy density and $m$ is the (light) quark mass (hereby I make no distinction between the mass of $u$ and $d$-quarks).

In the chiral limit ($m=0$) the quark condensate is a true order parameter since it is finite in the chirally broken (Nambu-Goldstone) phase and vanishes in the chirally symmetric (Wigner-Weyl) phase. In this limit the (true) critical temperature can be obtained from the {\it chiral susceptibility} 
\bea
   \chi_m=\frac{\partial}{\partial m}\langle\bar{\psi}\psi\rangle,
\label{susceptibility}
\eea
which has the critical behavior $\chi_m\sim m^{1/\delta - 1}$. For $O(N)$ scaling $\delta > 1$. Therefore $\chi_m$ diverges at the critical temperature, where $m=0$. 

For physical quark masses neither $\langle\bar{\psi}\psi\rangle$ vanishes nor $\chi_m$ diverges at the pseudocritical temperature. Nevertheless these quantities retain a reminiscent behavior of the corresponding one in the chiral limit. In particular $\chi_m$ has a peaked structure as a function of $T$ and it is customary to define $T_c$ as the temperature for which $\chi_m$ reaches its peak.  Other susceptibilities, such as $\chi_T=\frac{\partial}{\partial T}\langle\bar{\psi}\psi\rangle$, can also be used to define $T_c$. It has been shown~\cite{Sheng} that the critical temperatures thus obtained coincide within a narrow band and therefore using any of these susceptibilities yields basically the same $T_c$.

Lattice calculations have provided values for $T_c$ extracted from the peak of $\chi_m$ for 2+1 flavors using different types of improved staggered fermions~\cite{Levkova}. These values show some discrepancies. The MILC collaboration~\cite{MILC}  obtained $T_c=169(12)(4)$ MeV. The BNL-RBC-Bielefeld collaboration~\cite{BNL-RBC-Bielefeld} reported $T_c=193(7)(4)$ MeV. The Wuppertal-Budapest collaboration~\cite{Wuppertal-Budapest} has consistently obtained smaller values, the latest being $T_c=147(2)(3)$ MeV. The HotQCD collaboration~\cite{HotQCD} has reported $T_c=154(9)$ MeV. The differences could perhaps be attributed to different lattice spacings.

To describe deconfinement a variable that is often used is the Polyakov loop $L,$ which is regarded as an order parameter in such a way that the thermal average of its trace, $\langle{\mbox{Tr}}\ L\rangle$, vanishes in the confined phase and is finite in the deconfined phase. The Polyakov loop is a particular kind of Wilson loop where the integration path $C$ is taken along the periodic time direction, namely
\bea
\label{Polyakov}
   L={\mathcal{P}}e^{ig\oint_Cdx^\mu A_\mu},
\eea
where ${\mathcal{P}}$ is the path order operator, $g$ the QCD coupling constant and $A_\mu$ is the gauge field. It can be shown~\cite{Greensite} that $\langle {\mbox{Tr}}\ L\rangle$ satisfies
\bea
\label{satisfies}
   \langle{\mbox{Tr}}\ L\rangle \propto e^{-\Delta F_q/T},
\eea
where $\Delta F_q$ is the difference between the free-energy of the gauge theory containing an isolated static quark and the free-energy of a pure gauge system. Equation~(\ref{satisfies}) means that  $\langle{\mbox{Tr}}\ L\rangle=0$ if and only if, for a given $T$, $\Delta F_q$ becomes infinite. This is interpreted as the impossibility to have an isolated quark and therefore this situation must correspond to the confined case. However, when $\langle{\mbox{Tr}}\ L\rangle\neq 0$ for a given $T$, $\Delta F_q$ must be finite. This means that quarks are deconfined, in the sense that they exist as isolated, finite energy states bearing their corresponding quantum numbers. 

An isolated static quark can be regarded as an infinitely heavy one. Therefore, for physical quark masses the Polyakov loop is no longer an order parameter and the extraction of $T_c$ is based on less firm soil. Nevertheless, lattice calculations have also turned their attention to the computation of the Polyakov loop. The HotQCD collaboration has made systematic studies showing that, once discretization effects are accounted for, the Polyakov loop yields consistent estimates of the thermal properties of the theory with a small quark mass dependence in the transition region. The Polyakov loop however rises faster for heavier quarks at higher temperatures~\cite{HotQCD}.  

The unambiguous picture presented by lattice QCD for $T\geq 0$, $\mu=0$ cannot be easily extended to the case $\mu\neq 0$, the reason being that standard Monte Carlo simulations can only be applied to the case where $\mu$ is either 0 or purely imaginary. Simulations with $\mu\neq 0$ are hindered by the {\it sign problem}~\cite{Forcrand}. Recall that in the computation of the QCD partition function with finite $\mu$, integration over each fermion field produces a {\it fermion determinant}, i.e. a factor ${\mbox{Det}}M\equiv{\mbox{Det}}(\slshh{D} + m +\mu\gamma_0)$, where $M$ is the fermion matrix. Let us consider in general a complex $\mu$. Taking the determinant on both sides of the identity
\bea
   \gamma_5(\slshh{D} + m +\mu\gamma_0)\gamma_5 = (\slshh{D} + m - \mu^*\gamma_0)^\dagger,
\label{gamma5identity}
\eea
we get
\bea
   {\mbox{Det}}(\slshh{D} + m +\mu\gamma_0)=\left[{\mbox{Det}}(\slshh{D} + m -\mu^*\gamma_0)\right]^*,
\label{detcomplex}
\eea
which shows that the fermion determinant is not real unless $\mu=0$ or purely imaginary. For real $\mu$ the direct sampling of a finite density ensemble by Monte Carlo methods is not possible, since the sampling requires a real non-negative measure. This problem is referred to as the sign problem, although a more appropriate name would seem to be the {\it phase problem}.

That the integrand of the partition function becomes complex would seem to be only a minor inconvenience.  A naive approach to deal with the sign problem would be to write~\cite{Aarts}
\bea
   {\mbox{Det}}M=|{\mbox{Det}}M|e^{i\theta}.
\label{simple}
\eea
To compute the thermal average of an observable $O$ in QCD one writes
\bea
   \langle O\rangle&=&\frac{\int DUe^{S_{\small{YM}}}{\mbox{Det}}M\ O}
   {\int DUe^{S_{\small{YM}}}{\mbox{Det}}M}
   \ = \ \frac{\int DUe^{S_{\small{YM}}}|{\mbox{Det}}M|e^{i\theta}\ O}
   {\int DUe^{S_{\small{YM}}}|{\mbox{Det}}M|e^{i\theta}},
\label{average}
\eea
where $S_{\small{YM}}$ is the Yang-Mills action. Notice that written in this manner, the simulations could be implemented in terms of the phase-quenched (pq) theory where the measure involves $|{\mbox{Det}}M|$ and the thermal average could be written as
\bea
   \langle O\rangle=\frac{\langle Oe^{i\theta}\rangle_{\mbox{\small{pq}}}}
   {\langle e^{i\theta}\rangle_{\mbox{\small{pq}}}}.
\label{pq}
\eea
The average phase factor (also referred to as the average sign) in the phase-quenched theory can be written as
\bea
   \langle e^{i\theta}\rangle_{\mbox{\small{pq}}}=e^{-V(f-f_{pq})/T},
\label{sign}
\eea
where $f$ and $f_{pq}$ are the free-energy densities in the full and phase-quenched theories, respectively and $V$ is the three-dimensional volume. If $f-f_{pq}\neq 0$, the average phase factor decreases exponentially when $V$ goes to infinity (the thermodynamic limit) and/or $T$ goes to zero. Under these circumstances the signal to noise ratio worsens. This is referred to as the {\it severe sign problem}.

To circumvent the sign problem, a possibility is to determine the first Taylor coefficients in the expansion of a given observable in powers of $\mu_B/T$. The coefficients of the expansion can be expressed as expectation values of traces of matrix polynomials taken in the $\mu_B=0$ ensemble.  Although care has to be taken regarding the growth of the statistical errors with the order of the expansion, this strategy has provided a very important result: the curvature of the transition line at $\mu_B=0$. The curvature $\kappa$ is defined as the dimensionless coefficient of the quadratic term in the Taylor expansion of the pseudocritical line  
\bea
   T_c(\mu_B)&=&T_c\left(1 - \kappa\frac{\mu_B^2}{T_c^2}\right)\;,\qquad 
   \kappa\ \equiv\ -\left.\left(T_c\frac{dT_c(\mu_B)}{d\mu_B^2}\right)\right|_{\mu_B=0}.
\label{curvature}
\eea
Values for $\kappa$ between 0.006 -- 0.02 have been reported~\cite{lattcurv}. It should be noted that since the phases are separated by a crossover, the curvature should depend in principle on the observable considered. Nevertheless, these curvatures give considerable smaller values than that of the chemical freeze-out curve obtained from statistical models~\cite{Cleymans}. This observation could be of potential importance for if the pseudo critical line is flatter than the chemical freeze-out line, the distance between the possible QCD critical end point (CEP) and the freeze-out curve increases. If this happens then possible experimental evidences for criticality may be washed out by the moment when particle abundances are established after a heavy-ion collision.

To distinguish between chiral and deconfinement transitions it has been suggested that one can look at the effects of a magnetic background, which enters as a new control parameter for the thermodynamics. Peripheral heavy-ion collisions might be used as a physically realizable situation to explore this possibility since they generate very intense magnetic fields, even stronger than magnetars. The possible splitting of the deconfinement and chiral transitions in an external magnetic field has been studied in Ref.~\cite{Mizher} within the linear sigma model coupled to quarks (with two flavors) and to Polyakov loops. The authors found that the vacuum correction from quarks on the phase structure was dramatic. When ignoring this correction, the confinement and chiral phase transition lines coincide.  However, inclusion of the correction led to an splitting of the confinement and chiral transition lines, and both chiral and deconfining critical temperatures became increasing functions of the magnetic field. The vacuum contribution from the quarks drastically affected the chiral sector as well. Since these conclusions were drawn from a model that does not reproduce the behavior of the critical temperature as a function of the magnetic field, found by lattice QCD calculations, they are nowadays regarded as incorrect. Furthermore, a recent lattice result shows that no significant difference between chiral and deconfinement transition temperatures exists up to fields as intense as 3.25 GeV$^2$~\cite{Endrodimag}. On the contrary, the different definitions of $T_c$ tend to approach each other and they cluster between 109 - 112 MeV at this high value of the magnetic field.

The behavior of strongly interacting matter in the presence of magnetic fields has become a subject of increasing interest over the past few years. This interest has been sparked by recent lattice results showing that the transition temperature with 2 + 1 quark flavors decreases with increasing magnetic field and that the strength of the condensate decreases for temperatures close and above the phase transition~\cite{Fodor,Bali2,Bali3}. This behavior has been dubbed {\it inverse magnetic catalysis}, and it has been the center of attention in a large number of model-dependent analyses~\cite{Noronha} --~\cite{Ferreira1}. For recent reviews see Refs.~\cite{Andersenreview, Miransky}. In general terms, it seems that inverse magnetic catalysis is not obtained in mean field approaches describing the thermal environment~\cite{Mizher},~\cite{Andersen} --~\cite{Fraga2}, nor when calculations beyond mean field do not include magnetic effects on the coupling constants~\cite{ahmrv}.

The novel feature implemented in effective models, able to account self-consistently for inverse magnetic catalysis, is the decrease of the coupling constants with increasing field strength obtained from the model itself~\cite{amlz} without resorting to {\it ad hoc} parametrizations. This has been achieved within the Abelian Higgs model and the linear sigma model coupled to quarks (LSMq). This behavior is made possible by accounting for the screening properties of the plasma, which have been recently formulated consistently for theories with spontaneous symmetry breaking. This results in a formalism beyond the mean field approximation. Screening is also important to obtain a decrease of the QCD coupling constant with the magnetic field in the Hard Thermal Loop approximation~\cite{Ayala3}. In the opposite limit, it has also been shown by means of a QCD sum rules analysis at zero temperature, that both, the threshold energy for the onset of the continuum in the quark vector current spectral density (a phenomenological parameter that signals the onset of deconfinement) and the gluon condensate increase with increasing magnetic field, as expected~\cite{adhlrv}.

The LSMq has been also used to explore the phase diagram without magnetic fields~\cite{mich}. It was found that there are values for the model couplings that allow locating a critical end point (CEP) in the region where lattice inspired calculations find it~\cite{sharma}. Since the LSMq does not exhibit confinement, this behavior is attributed to the proper treatment of plasma screening, instead of to the existence of a given confinement length scale~\cite{roberts}. In addition, recent LQCD calculations~\cite{Endrodimag} show that for very strong magnetic fields, inverse magnetic catalysis prevails and the phase transition becomes first order at asymptotically large values of the magnetic field for vanishing quark chemical potential $\mu$. A similar behavior is obtained in the Nambu Jona-Lasinio model if one includes a magnetic field dependence of the critical temperature in agreement with lattice QCD results~\cite{Costa}.

In Ref.~\cite{magnetized} the LSMq was used to explore the consequences of a proper handling of the plasma screening properties in the description of the magnetized effective QCD phase diagram. It was shown that when including self-consistently magnetic field effects in the calculation of both the effective potential as well as on the thermo-magnetic dependence of the coupling constants, the CEP moves toward smaller values of the critical quark chemical potential and larger values of the critical temperature. In addition, above a certain value of the field strength the CEP moves toward the $T$-axis. This behavior can be understood on general grounds, as the magnetic field produces a dimensional reduction, whereby virtual charged particles from the vacuum are effectively constrained to occupy Landau levels, thus restricting their motion to a plane. This makes these particles to lay closer to each other on average, thus reducing the interaction strength for strongly coupled theories. This situation takes place regardless of how weak the external field is. 

The search for the CEP is one of the main motivations for the RHIC beam energy scan (BES) program. The two RHIC experiments, PHENIX and STAR, have reported their results for the BEC phase I (see for example Refs.~\cite{PHENIXBEC} and~\cite{STARBEC}) in Au + Au collisions at $\sqrt{s_{NN}}=7.7, 11.5, 19.6, 27, 39$, and 62.4 GeV. As the collision energy decreases, the stopping power becomes larger and therefore $\mu_B$ becomes also larger which in turn allows to experimentally probe deeper into the phase diagram. The experimental findings reveal that for central collisions, as the collision energy lowers, several signals of the QGP fade away. Among them we can mention the disappearance of the elliptic flow scaling with the number of constituent quarks, the growth in the difference of the elliptic flow between particles and anti-particles~\cite{STAR-v} and the smaller strength of the pion $R_{AA}$~\cite{PHENIX-pi0}. 

Of particular relevance for identifying and locating the CEP is the search for event-by-event fluctuations of conserved charges. The idea is that these fluctuations are proportional to powers of the relevant correlation length which in turn must diverge, or at least present a rapid growth, near criticality. The analogue is the liquid-gas transition where the {\it critical opalescence} phenomenon takes place as the critical point is approached. In this region of the phase diagram, the sizes of the gas and liquid domains begin to fluctuate over larger length scales. As the size of the density fluctuations (the correlation length) becomes comparable to the wavelength of light, this last is scattered in all directions causing the normally transparent liquid to appear cloudy, or opal-like, and hence the name of the phenomenon. 

It has been recently argued that in order to look for critical behavior it is important to target moments of higher order than quadratic of the event-by-event conserved charge's distribution, since these are proportional to higher powers of the correlation length and thus in principle easier to detect~\cite{Stephanov}. It is even better to look for ratios of these moments (or of the cumulants)~\cite{Gupta1} since parameters such as the fireball volume, which are poorly known, cancel and the ratios are better comparable to lattice predictions~\cite{Bazavov2}. The shapes of the distributions are also expected to be sensitive to the presence of the critical point~\cite{Gavai}. PHENIX has measured the skewness and the kurtosis of net charge distributions for four different collision energies $\sqrt{s_{NN}}=200, 62.4, 39$, and 7.7 GeV finding no excess above the values from simulations that do not include effects of critical behavior. In contrast, the CERN-NA49 collaboration has reported a significant intermittency signal in the  transverse momentum correlations of opposite sign pions in Si + Si collisions at top SPS energies~\cite{NA49} approaching in size some predictions for critical behavior in QCD~\cite{Antoniou}. 
 
The experimental search for the structure of the phase diagram for hot and dense matter will continue with the RHIC-BES phase II scheduled to start in 2018 and for cold nuclear matter with the starting of the EIC program sometime after 2024. The NA61/SHINE beam energy and system size scan, started in 2009, will continue pursuing evidence for the existence of the critical point hinted by NA49 fluctuation data~\cite{Gazdzicki}. In addition new planned facilities like the Facility for Antiprotons and Ion Research (FAIR) in GSI and the Heavy-Ion Collider Facility NICA at JINR-Dubna are under construction and promise to deepen the exploration of the QCD phase diagram in the future. 

\section{Analytical signals of chiral symmetry restoration/deconfinement}\label{ii}

The connection between parameters describing chiral symmetry restoration/deconfinement and observables such as the hadron spectral density is provided by the QCD sum rules (QCDSRs) approach. This analytical method is, to a large extent, complementary to the numerical lattice simulations of hadrons. In this section we concentrate on summarizing recent progress employing the QCDSRs approach for the computation of the hadronic spectral density.

The QCDSRs method was developed more than thirty years ago~\cite{sumrule} and ever since it has become a prime working tool in hadron phenomenlogy. QCDSRs rest on two pillars~\cite{Colangelo}, (i) the operator product expansion (OPE) of current correlators at short distances beyond perturbation theory, and (ii) Cauchy's theorem in the complex squared energy $s$-plane (a dispersion integral), which relates the (hadronic) discontinuity across the cut on the real positive semi-axis with the integral around a contour of radius $|s_0|$ (which could be taken to infinity) where the OPE is expected to be valid. The latter is usually referred to as quark-hadron duality. 

An important object in QCD is the amplitude for the quark-pair creation and annihilation, the so called current correlator. The formal expression for this amplitude at $T,\mu=0$ can be written as
\bea
\label{correlator}
   \Pi_{\mu\nu}(q)=i\int d^4xe^{iq\cdot x}\langle 0 |{\mathcal{T}}[j_\mu (x)j_\nu^\dagger (0)]|0\rangle,
\eea
where $q^\mu=(q_0,{\mathbf{q}})$ is the four-momentum carried by the current, $j_\mu=\bar{\psi}\gamma_\mu\psi$ is the colorless quark-current of a given flavor (for our purposes $\psi=u,d,s,c\ldots$ or a linear combination of these), $\langle 0|\ldots |0\rangle$ is the vacuum expectation value and ${\mathcal{T}}$ is the time order product operator. Current conservation dictates that the Lorentz structure of the correlator is
\bea
\label{lorentzstructure}
   \Pi_{\mu\nu}(q)=-q^2\left(g_{\mu\nu}-\frac{q_\mu q_\nu}{q^2}\right)\Pi(q^2).
\eea
$\Pi(q^2\equiv s)$ contains all the dynamical effects.

At finite temperature and/or density the correlator can be written as
\bea
   \Pi_{\mu\nu} (q_0^2,{\mathbf{q}}^2)&=&i\int d^4x e^{iq\cdot x} \langle {\mathcal{T}}[j_\mu(x)j_\nu^\dagger (0)]
   \rangle\nn
   &=&-q^2\left[\Pi_0(q_0^2,{\mathbf{q}}^2)P^T_{\mu\nu}+\Pi_1(q_0^2,{\mathbf{q}}^2)P^L_{\mu\nu}\right],
\label{correlatorT}
\eea
where now $\langle\cdots\rangle$ refers to a thermal average and $P^T_{\mu\nu}$ and $P^L_{\mu\nu}$ are the three-dimensional transverse and longitudinal projectors (both are four dimensionally transverse), respectively. At finite temperature and/or density it is customary to work in the {\it static} limit ${\mathbf{q}}\rightarrow 0$. The rationale is that in equilibrium and for temperatures around the (pseudo)critical temperature, the average three-momentum is small compared to the binding energy (mass) of the relevant hadrons and therefore these last can be considered as slowly moving. In this limit $\Pi_{\mu\nu}$ contains only spatial components and therefore the only remaining coefficient in Eq.~(\ref{correlatorT}) is $\Pi_0(q_0^2\equiv s)$ which now carries the information on the dynamical effects. The temperature behavior of the light-quark vector current correlator in the framework of thermal QCDSRs was first discussed in Ref.~\cite{Bochkarev} and later reanalyzed by others~\cite{others}.

Let us argue how the spectral density is related to the physics of hadrons. $\Pi(s)$ and $\Pi_0(s)$ can be thought of as analytical functions of $s$ defined for negative (space-like), positive (time-like) or even complex values of $s$. Also, for large positive $s$, the integral in Eqs.~(\ref{correlator}) and~(\ref{correlatorT}) are dominated by large values of $x^2$~\cite{Colangelo}. This means that the long distance QCD interactions become important, forcing the quarks in the current to eventually form hadrons. In particular, a quark-antiquark pair created by the current $j_\mu$ with $J^P=1^-$ becomes a neutral vector meson. The ground-state vector mesons with the isospin $I = 1$ and $I = 0$ are $\rho$ and $\omega$ having the quark content
$(\bar{u}u - \bar{d}d)/\sqrt{2}$ and $(\bar{u}u + \bar{d}d)/\sqrt{2}$, respectively. The lightest vector meson
created by the current $\bar{s}\gamma_\mu s$ is $\phi$. For large negative values of $s$, the integral in Eqs.~(\ref{correlator}) and~(\ref{correlatorT}) are dominated by small values of $x^2$. Therefore quarks propagate predominantly at short distances and during short time intervals. The QCD interactions are suppressed due to asymptotic freedom. Therefore, as a first approximation, one may calculate the current correlator using perturbation theory.  At finite temperature and/or density, one needs to restrict this perturbative piece to the leading, one-loop level, as the treatment of the appearance of two scales in $\alpha_s (q^2,T)$, i.e., $\Lambda_{\small{QCD}}$ and $T_c$, remains  an open problem. Corrections to this approximation should account for the fact that the QCD vacuum influences this propagation. This influence is parametrized in terms of the condensates and expressed in terms of the OPE. 

A direct measurement of the in-medium hadronic spectral density in the vector channel, $\Pi_V$, is provided by the invariant-mass spectrum of low mass dileptons. Dileptons and photons constitute what is generically called {\it electromagnetic probes} in heavy-ion collisions. These probes are emitted continuously from the early quark-gluon plasma phase through to the late hadronic phase, therefore revealing the entire thermal evolution of the reaction. Since their mean free path is larger than the size of the fireball, these probes escape from the medium without further interaction. For invariant masses below 1 GeV, the spectrum is dominated by the light vector mesons $\rho$, $\omega$ and $\phi$. The temperature and density dependence of the spectral function for these particles reflects their mass distribution and thus opens a window to experimentally study the dynamical mass generation mechanism in QCD. Among these vectors, $\rho$ is an ideal test particle to sample in-medium changes of parameters such as mass, width and leptonic decay constant, given its short lifetime and large coupling to pions and muons. There is no similar experimental knowledge of the in-medium axial vector spectral density, $\Pi_A$.

Data from the CERN-NA60 Collaboration~\cite{NA60-1,NA60-2} has allowed to settle a long standing controversy regarding the origin of the dilepton excess~\cite{CERES} below the $\rho$ peak at SPS energies. The data shows that when the known sources are subtracted from the spectral function, a clear peak that broadens for the most central collisions but remains centered at its vacuum mass value is observed at all centralities. The total dimuon yield also increases with centrality. This result is in line with current ideas about the nature of the deconfinement/chiral symmetry restoration transition; a width which increases with increasing $T$ and/or $\mu$ indicates deconfinement, with the spectral function becoming smooth and eventually accounted for by perturbative QCD, irrespective of the changes in the mass~\cite{Dominguez1}. One could picture the extreme situation of the mass decreasing and
vanishing at some critical temperature. If this behavior is not accompanied by an increasing width then the hadron would still be present in the spectral function at that temperature.

The leading theoretical descriptions of the $\rho$ broadening at SPS energies are based on hadronic many-body or transport calculations, invoking the idea that the $\rho$ scatters and thus melts not only at finite temperature, but also in a baryon-rich environment~\cite{Rapp} --~\cite{Rapp2} (see also Ref.~\cite{Sarkar}). These rather successful descriptions have allowed the interpretation of a wealth of fixed target data. In principle, this approach could also be expected to describe data at RHIC and LHC, where the net density remains comparable to SPS energies although the baryon density is smaller. However, the PHENIX collaboration has found a large enhancement in central Au + Au at $\sqrt{s_{NN}}= 200$ GeV~\cite{PHENIX} which cannot be described by the in-medium hadronic effects invoked at SPS energies, albeit this description works well for lower centralities and lower energies. This enhancement is not reported by STAR~\cite{STAR-dielectrons}. Recently, PHENIX has reported new measurements with a better signal to background ratio performed with their Hadron Blind Detector which come closer to STAR results~\cite{PHENIX2}.

One way to connect the spectral densities $\Pi_V$ and $\Pi_A$ with a signal of chiral symmetry restoration is through a possible mixing of their vacuum values $\Pi_V^0$ and $\Pi_A^0$. This idea was first pursued in Ref.~\cite{Dey} and led to the deduction of the expression
\bea
   \Pi_{V,A}(q^2;T)=[1-\epsilon (T)]\Pi_{V,A}^0(q^2) + \epsilon (T)\Pi_{A,V}^0(q^2).
\label{mixing}
\eea
Equation~(\ref{mixing}) is valid for low temperatures and when considering only pions in the sum over excited states in the Gibbs averages. It represents the leading $T$-dependence of the spectral densities which are related by the mixing parameter $\epsilon (T)$ which in the chiral limit, $m_\pi=0$, becomes $\epsilon (T)=T^2/6f_\pi^2$, where $f_\pi$ is the pion decay constant. The mixing parameter is proportional to the scalar pion density. Equation~(\ref{mixing}) satisfies Weinberg-type sum rules~\cite{Weinberg, Das} (WSR) provided $\Pi_{V,A}^0$ also satisfy them. The physical process responsible for the mixing is the interaction of $\rho$ with pions to form $a_1$. In the chiral limit, when $\epsilon (T)\simeq 0.5$, Eq.~(\ref{mixing}) shows that $\Pi_V$ and $\Pi_A$ become degenerate, which is interpreted as signaling chiral symmetry restoration. Notice however that even in the chiral limit, this happens for $T\simeq 160$ MeV, temperature for which other than only pion degrees of freedom are expected to be present and the above expression is no longer reliable.

When the mixing scenario is pursued, it has been argued that it is possible to use Eq.~(\ref{mixing}) to extract information on the relation between mixing of spectral densities and chiral symmetry restoration in a more sophisticated setup. This has been done in Ref.~\cite{Holt} where use is made of vacuum spectral densities containing a postulated universal separation of nonperturbative and perturbative regimes. The former is represented by excited states and the latter by a degenerate continuum. The description includes also a finite pion mass. However, it is found that deviations from the QCDSRs set in at temperatures of the order of the pion mass suggesting the need of additional physics, beyond low-energy chiral pion dynamics, that is, the inclusion of resonances. This result may also indicate that the encoding of the finite temperature effects in such a simple mixing scenario may not be a good approximation for not too low temperatures.

More recently, another connection between chiral symmetry restoration/deconfinement and hadron spectral densities has been explored using the Finite Energy QCDSRs (FESR) approach. The method involves relating the moments of the hadronic spectral densities to condensates. However, the integrals are computed up to a {\it finite squared-energy} $s_0$, and thus the name of the method. $s_0$ corresponds to the square of the energy threshold for the onset of the continuum.  This is a key parameter signaling deconfinement above which the hadronic spectral function is well approximated by perturbative QCD. The importance of this parameter was first pointed out in Ref.~\cite{Bochkarev}. Explicit determinations of $s_0$ in several light and heavy-light quark systems~\cite{others},~\cite{Dominguez3} --~\cite{Dominguez6} find this parameter to decrease with increasing temperature, vanishing at a critical value interpreted as the deconfinement temperature. 

Integrating the function $(s^N/\pi)/\Pi_0(s)$ on Eq.~(\ref{correlatorT}) in the complex $s$-plane along a contour with a fixed radius $|s| = s_0$, deforming the contour to make it pass slightly below and above the the positive real axis and assuming that $\Pi_0(s)$ is analytical within the integration contour we obtain the FESR
\bea
   (-1)^{N-1}C_{2N}\langle O_{2N}\rangle = 8\pi^2\left[
   \int_0^{s_0}dss^{N-1}\frac{1}{\pi}{\mbox{Im}}\Pi^{\mbox{\small{HAD}}}(s)
   -
   \int_0^{s_0}dss^{N-1}\frac{1}{\pi}{\mbox{Im}}\Pi^{\mbox{\small{pQCD}}}(s)\right],
\label{FESR}
\eea
where $N =1,\ 2,\ldots$, and the leading order vacuum condensates in the chiral limit are the dimension $d = 4$ gluon condensate $C_4\langle O_4\rangle = (\pi/3)\langle\alpha_sG^2\rangle$, and the dimension $d=6$ four-quark condensate, $C_6\langle O_6\rangle$. The implicit assumption is that $s_0(T,\mu)$ is large enough as to preserve the validity of the OPE with varying temperature and/or chemical potential. 

An interesting connection between $s_0(T,\mu)$ and $\langle\bar{\psi}\psi\rangle(T,\mu)$ has been found in Ref.~\cite{morelia} in the analysis of the FESR in the axial channel. Results indicate that the critical temperatures for deconfinement and chiral symmetry restoration are basically the same within a few percent accuracy up to about $\mu=100$ MeV where the method is reliable. FESR in the vector channel have also been used to determine the parameters describing the $\rho$-meson resonance.~\cite{nos}. Assuming $\rho$ saturation of
the spectral function, and a Breit-Wigner resonance form
\bea
   \frac{1}{\pi}{\mbox{Im}}^{\mbox{\small{HAD}}}(s)=\frac{1}{\pi}
   \frac{1}{f_\rho^2}\frac{M_\rho^3\Gamma_\rho}{(s-M_\rho^2)^2+M_\rho^2\Gamma_\rho^2},
\label{BWform}
\eea
it is found that a consistent solution for the three leading FESR are obtained when the $\rho$-width, $\Gamma_\rho$, increases and even diverges at $T_c$ whereas the $\rho$-mass, $M_\rho$, and the leptonic decay constant, $f_\rho$, decrease only slightly near $T_c$. 

This solution has recently been extended to finite $\mu$ and used to describe the dilepton spectrum around the $\rho$-peak~\cite{Ayala2-2}. The extension to finite chemical potential is made by first including the $\mu$-dependence into the quark loop of the pQCD piece in Eq.~(\ref{FESR}) which splits the Fermi-Dirac distribution into particle--antiparticle contributions and then by incorporating the $\mu$-dependence of the critical temperature $T_c$. This dependence is obtained from the results of Ref.~\cite{morelia2} that make use of a Schwinger-Dyson approach, to find a parametrization for the crossover transition line between chiral symmetry restored and broken phases, valid for small values of $\mu$,
\bea
   T_c(\mu)=T_c(\mu=0) - 0.218\mu - 0.139\mu^2.
\label{tcvsmu}
\eea
Assuming a simple, Bjorken-like expansion cooling law, the results are in good agreement with NA60 dimuon data~\cite{NA60-2}. FESR have also been used in the axial-vector channel to explore the relation between the (pseudo)critical temperature for deconfinement transition --signaled by the vanishing of $s_0$-- and that for chiral symmetry restoration --signaled by the vanishing of the light-quark condensate~\cite{DLZ}. The result is that both temperatures coincide within the accuracy of the method.

The mixing scenario has been challenged in Ref.~\cite{NoMixing}. The argument is based on the dramatic difference between the vector and the axial-vector spectral functions already at $T=0$ and on their very different evolution with increasing temperature, with the exception that $s_0$ is the same in both channels. With increasing $T$, the parameters of each channel evolve independently, thus keeping both spectral functions distinct. Eventually, this asymmetry is expected to vanish at deconfinement, when the $\rho$ and the $a_1$ mesons disappear from the spectrum. This implies no chiral-mixing at any temperature, except obviously at $T\simeq T_c$. In addition to these differences there is an asymmetry due to the hadronic (pionic) scattering term, present in the vector channel at the leading, one-loop level, and strongly two-loop suppressed in the axial-vector case~\cite{ReviewADL}. To be more specific let us consider the vector and axial-vector correlators. In a thermal bath, and in the hadronic representation one has (schematically)
\bea
    	\Pi_{\mu\nu}|_V &=& \langle\pi | V_\mu (0)V_\nu(x)  | \pi\rangle = 
	\langle\pi | V_\mu (0) | \pi\rangle \langle\pi | V_\nu(x) | \pi\rangle +
	\langle\pi\pi | V_\mu (0) | \pi\pi\rangle \langle\pi\pi | V_\nu(x) | \pi\pi\rangle + \ldots\nn
	\Pi_{\mu\nu}|_A &=&\langle\pi | A_\mu (0)A_\nu(x)  | \pi\rangle = 
	\langle\pi | A_\mu (0) | 0\rangle \langle 0 | A_\nu(x) | \pi\rangle +
	\langle\pi\pi\pi | A_\mu (0) | 0\rangle \langle 0 | A_\nu(x) | \pi\pi\pi\rangle + \ldots\nn
\label{axial-vector-corr}
\eea
Notice that in Eq.~(\ref{axial-vector-corr}) the chiral asymmetry is manifest at finite $T$, since Isospin and $G$-parity remain valid symmetries also at finite $T$. The leading term in the vector channel is the two-pion loop whereas in the axial-vector channel the leading term is the tree-level pion to-vacuum term $(f_\pi)$ (pion pole) followed by a highly phase-space suppressed three-pion loop. In other words, the matrix element $\langle\pi | A_\mu (0) | 0\rangle$, needed for chiral-mixing, vanishes identically at leading order. The correct matrix element, beyond the pion pole, is the phase-space suppressed second term on the second of Eqs.~(\ref{axial-vector-corr}). In principle, this term could have a one-loop resonant sub-channel contribution from the matrix element $\langle\rho\pi | A_\mu (0) | 0\rangle$. However, this contribution is also phase space suppressed due to the finite (albeit not too large) $\rho$-width. 

Thermal Hilbert moment QCD sum rules have also been used to determine the behavior of hadronic parameters of charmonium in the scalar, pseudoscalar and vector channels at finite temperature~\cite{Dominguez7}. From the results of the temperature dependence of the width and the leptonic coupling, the charmonium channels show signs of survival beyond the deconfinement pseudo critical temperature $T_c$. It should be noted in contrast that the light-light and even the heavy-light quark mesons disappear at that $T_c$. The difference is driven by two main effects: (1) the $T$-dependence of the light-light and heavy-light quarks systems in the pQCD sector is dominated by the spectral function in the time-like region, the so-called annihilation term, which is anyway relatively unimportant in relation to the light quark condensate contribution. The pQCD spectral function in the space-like region (scattering term) is highly suppressed. For heavy-heavy quark systems the scattering term becomes increasingly important with increasing temperature, while the annihilation term only contributes near threshold. (2) in the nonperturbative QCD sector of light-light and heavy-light quark correlators, the driving term in the OPE is the light quark condensate. This term is responsible for the behavior of $s_0(T)$. In contrast, for heavy-heavy quark correlators the leading power correction in the OPE is that of the gluon condensate, which has a very different temperature behavior. 

The distinct signatures of the survival of charmonium states beyond $T_c$ are the behavior of the
width and the leptonic coupling. These parameters are almost independent of $T$ up to $T\simeq 0.8T_c$ where the width begins to increase substantially. However, above $T\simeq 1.04 T_c$ it starts to decrease sharply and the coupling increases also sharply. This behavior, which can mostly be traced back to that of the pQCD spectral function in the space-like region, points to the survival of the $J/\psi$, $\eta_c$ and $\chi_c$ resonances above the deconfinement temperature. However, the QCD sum rules have no longer solutions for the hadronic parameters once the continuum threshold reaches the value $s_0|_{\mbox{\tiny{min}}}\simeq M^2_{J/\psi}$, as there is no longer any support for the integrals. Hence, the temperature range explored with this technique does not extend beyond $T\simeq 1.22T_c$. A similar behavior is obtained for bottomonium ground states in the vector $(\Upsilon)$ and pseudo scalar $(\eta_b)$ channels~\cite{Dominguez8}. The width increases with temperature, but for $T/T_c\simeq 0.9$ it drops approaching its zero temperature value. The leptonic decay constant is basically a monotonically increasing function of $T$. The results are also interpreted as the survival
of these bottonium states above $T_c$.

The QCDSRs method is powerful and can be implemented in a wide range of forms. Although it has been applied to the study of physical systems at finite temperature and density ever since only a few years after its original formulation, the heavy-ion program has provided the motivation to continue exploring this tool and to relate it to other analytical and numerical methods such as SDEs and lattice studies. The method promises to continue presenting insightful ways to look at the physical systems under extreme conditions of temperature and density.

\section*{Acknowledgments}

This work has been supported in part  by  UNAM-DGAPA-PAPIIT grant number IN101515 and by Consejo Nacional de Ciencia y Tecnolog\1a grant number 256494.

\end{document}